\documentclass[a4]{article} 
\makeatletter
\renewcommand{\section}{\@startsection{section}{1}{0pt}{-\baselineskip}%
{0.5\baselineskip}{\large\normalfont\bfseries}}
\newcommand{\DoS}{{\mathcal{S}}}
\newcommand{\row}[1]%
{\mathord{\buildrel{\lower3pt%
\hbox{$\scriptscriptstyle\rightarrow$}}\over #1}}
\newcommand{\col}[1]{{#1^{\raisebox{2pt}[\height]%
{$\scriptstyle\downarrow$}}}}
\newcommand{\dyadic}[1]{\mathord{\dyadic@rrow{#1}}}
\newcommand{\dyadic@rrow}[1]{
\begin{picture}(12,12)(-1,0)
\put(0,11){\makebox(0,0)[t]{$\scriptscriptstyle\downarrow$}}
\put(-1,10.5){\makebox(0,0)[l]{$\scriptscriptstyle\longrightarrow$}}
\put(5,0){\makebox(0,0)[b]{$#1$}}
\end{picture}
}
\newcommand{\trans}[1]{{#1}^{\raisebox{-1pt}{\scriptsize\textrm{T}}}}

\newcommand{\Tr}[1]{{\mathrm{Tr}}\left\{#1\right\}}
\newcommand{\Spur}[1]{\mathop{\mathrm{Sp}\left\{#1\right\}}}
\newcommand{\determ}[1]{\mathop{\mathrm{det}\left\{#1\right\}}}
\newcommand{\Rho}{{\mathrm{P}}}
\newcommand{\tRho}{\widetilde{\mathrm{P}}}
\newcommand{\tK}{\widetilde{K}}
\newcommand{\tkappa}{\widetilde{\kappa}}
\newcommand{\bRho}{\widebar{\mathrm{P}}}
\newcommand{\Bell}{\Rho_{\rm{Bell}}}
\newcommand{\chaos}{\Rho_{\rm{chaos}}}
\newcommand{\pure}{\Rho_{\rm{pure}}}
\newcommand{\tpure}{\widetilde{\Rho}_{\rm{pure}}}
\newcommand{\sep}{\Rho_{\rm sep}}
\newcommand{\pureopt}{\Rho^{\rm(opt)}_{\rm{pure}}}
\newcommand{\sepopt}{\Rho^{\rm(opt)}_{\rm{sep}}}
\newcommand{\tsepopt}{\tRho^{\rm(opt)}_{\rm{sep}}}
\newcommand{\widebar}[1]{\overline{#1}}
\newcommand{\expect}[1]{\left\langle #1 \right\rangle}
\newcommand{\olexpect}[1]{\bigl\langle #1 \bigr\rangle}
\newcommand{\ROW}[1]{\left( #1 \right)}
\newcommand{\COL}[1]{\left(\begin{array}{c} #1 \end{array}\right)}
\newcommand{\DYD}[1]{\left(\begin{array}{ccc} #1 \end{array}\right)}
\newcommand{\half}{\frac{1}{2}}
\newcommand{\AND}{\quad\mbox{and}\quad}
\newcommand{\OR}{\quad\mbox{or}\quad}
\newcommand{\WITH}{\quad\mbox{with}\quad}
\newcommand{\PHHH}{PH$_3$}\newcommand{\HW}{HW}
\newcommand{\WHC}{{\mathcal{C}}}
\newcommand{\magn}[1]{%
{\mathchoice{\magn@D{#1}}{\magn@T{#1}}{\magn@S{#1}}{\magn@SS{#1}}}}
\newlength{\@xx}\newlength{\@yy}\newlength{\@zz}
\newcommand{\magn@D}[1]{%
\settoheight{\@xx}{$\displaystyle\left|#1\right|$}%
\settodepth{\@yy}{$\displaystyle\left|#1\right|$}%
\addtolength{\@xx}{\@yy}%
{\,\rule[-\@yy]{\@zz}{\@xx}\,#1\,\rule[-\@yy]{\@zz}{\@xx}\,}}
\newcommand{\magn@T}[1]{%
\settoheight{\@xx}{$\textstyle\left|#1\right|$}%
\settodepth{\@yy}{$\textstyle\left|#1\right|$}%
\addtolength{\@xx}{\@yy}%
{\,\rule[-\@yy]{\@zz}{\@xx}\,#1\,\rule[-\@yy]{\@zz}{\@xx}\,}}
\newcommand{\magn@S}[1]{%
\settoheight{\@xx}{$\scriptstyle\left|#1\right|$}%
\settodepth{\@yy}{$\scriptstyle\left|#1\right|$}%
\addtolength{\@xx}{\@yy}%
{\,\rule[-\@yy]{\@zz}{\@xx}\,#1\,\rule[-\@yy]{\@zz}{\@xx}\,}}
\newcommand{\magn@SS}[1]{%
\settoheight{\@xx}{$\scriptscriptstyle\left|#1\right|$}%
\settodepth{\@yy}{$\scriptscriptstyle\left|#1\right|$}%
\addtolength{\@xx}{\@yy}%
{\,\rule[-\@yy]{\@zz}{\@xx}\,#1\,\rule[-\@yy]{\@zz}{\@xx}\,}}
\makeatother

\begin{document}

\title{\mbox{\Large\textbf{Remarks on 2--q-bit states}}}
\author{%
Berthold-Georg Englert$^{1,2}$ and Nasser Metwally$^3$\\ 
\parbox{160pt}{\rule{0pt}{14pt}\small\makebox[0pt][r]{$^1$}%
Max-Planck-Institut f\"ur Quantenoptik\\
Hans-Kopfermann-Strasse 1\\ 85748 Garching, Germany}
\\ 
\parbox{160pt}{\rule{0pt}{14pt}\small\makebox[0pt][r]{$^2$}%
Abteilung f\"ur Quantenphysik\\ 
Universit\"at Ulm\\
Albert-Einstein-Allee~1\\ 
89081 Ulm, Germany}
\\ 
\parbox{160pt}{\rule{0pt}{14pt}\small\makebox[0pt][r]{$^3$}%
Sektion Physik\\ Universit\"at M\"unchen\\ 
Theresienstra\ss{}e 37\\ 80333 M\"unchen, Germany}
}

\date{17 July 2000}

\maketitle

\begin{abstract}%
We distinguish six classes of families of locally equivalent states
in a straightforward scheme for classifying all 2--q-bit states;
four of the classes consist of two subclasses each.
The simple criteria that we stated recently for checking a given state's
positivity and separability are justified, and we discuss some important
properties of Lewenstein-Sanpera decompositions.
An upper bound is conjectured for the sum of the degree of separability of a
2--q-bit state and its concurrence.
\end{abstract}

\section{Introduction}\label{sec:intro}
Entangled q-bits (binary quantum alternatives) are exploited
in most schemes proposed for quantum communication purposes, for
quantum information processing, or for the secure key distribution procedures
known as quantum cryptography.
The basic units are entangled q-bit pairs. 
Obviously then, a thorough understanding of the properties of 2--q-bit states
is desirable.
Although there has been considerable progress in this matter recently, the
situation is still quite unsatisfactory.

The characterization of the 2--q-bit states produced by some source requires
the experimental determination of 15 real parameters.
Ideally, this is done by measuring a suitably chosen set of five observables
\cite{WF} that constitute ``a complete set of five pairs of complementary
propositions'' \cite{BZ}.
In an optical model \cite{EngKuWei}, which makes use of single-photon 2--q-bit
states, these measurements can be realized, and other experimental studies of
2--q-bit states can be performed as well.

Then, based on the knowledge of the 15 state-specify\-ing parameters, one can
classify the 2--q-bit state.
We distinguish, in Sec.\ \ref{sec:families}, six classes of families of
locally equivalent states.
Roughly speaking, local equivalence means that the difference is of a
geometrical, not a physical nature.
In a certain sense, the 15 parameters can be regarded as consisting of 6
geometrical ones and 9 physical ones.

The classification of Sec.\ \ref{sec:families} is straightforward but not
sufficient.
One also needs to know if the 2--q-bit state in question is useful for quantum
communication purposes.
In the technical terms of Sec.\ \ref{sec:LSD}, we ask for its degree of
separability as a numerical measure for this usefulness.
The degree of separability is part and parcel of the so-called optimal 
Lewenstein-Sanpera decomposition \cite{LS} of a 2--q-bit state.
This decomposition is known for a number of relevant types of states
\cite{EngMet} but, despite a good understanding of its properties, presently
we do not have a method for finding it for any arbitrary 2--q-bit state.

In Sec.\ \ref{sec:concur} we remark briefly on the so-called concurrence of a
2--q-bit state \cite{HW,Woo} and surmise that the sum of the degree of
separability and the concurrence does not exceed unity.
The Appendix reports some technical details.

\section{Families of 2--q-bit states}
\label{sec:families}
We employ the terminology and the notational conventions of \cite{EngMet}.
As usual, we describe the individual q-bits with the aid of analogs of
Pauli's spin vector operator: $\row{\sigma}$ for the first q-bit, $\row{\tau}$
for the second.
These row vectors refer to two three-dimensional vector spaces that are
unrelated, which is to say that in
\begin{eqnarray}
\row{\sigma}&=&\sum_{\alpha=x,y,z}\sigma_{\alpha}\row{e_{\alpha}}
=\ROW{\sigma_x,\sigma_y,\sigma_z}\COL{\row{e_x} \\ \row{e_y} \\ \row{e_z}}
\,,\nonumber\\
\row{\tau}&=&\sum_{\beta=x,y,z}\tau_{\beta}\row{n_{\beta}}
=\ROW{\tau_x,\tau_y,\tau_z}\COL{\row{n_x} \\ \row{n_y} \\ \row{n_z}}
\label{eq:Pauli-ops,xyz}
\end{eqnarray}
the orthonormal right-handed vector sets $\row{e_x},\row{e_y},\row{e_z}$ and
$\row{n_x},\row{n_y},\row{n_z}$ have nothing to do with each other.

In addition to these pre-chosen $xyz$ coordinate systems, we'll also 
consider 123 coordinate systems that are adapted to the 2--q-bit state of
interest. 
Then
\begin{equation}\label{eq:Pauli-ops,123}
\row{\sigma}
=\ROW{\sigma_1,\sigma_2,\sigma_3}\COL{\row{e_1} \\ \row{e_2} \\ \row{e_3}}
\,,\quad
\row{\tau}
=\ROW{\tau_1,\tau_2,\tau_3}\COL{\row{n_1} \\ \row{n_2} \\ \row{n_3}}
\end{equation}
are the respective parameterizations of $\row{\sigma}$ and $\row{\tau}$.
As an elementary illustration think of the statistical operator of the 
first q-bit, specified by the Pauli vector $\row{s}=\olexpect{\row\sigma}$,
\begin{eqnarray}
  \rho_1=\half\bigl(1+\row{s}\cdot\col{\sigma}\bigr)
        &=&\half\bigl(1+s_x\sigma_x+s_y\sigma_y+s_z\sigma_z\bigr)
        \nonumber\\
        &=&\half\bigl(1+s\sigma_1)\,,
  \label{eq:rho1}
\end{eqnarray}
where $s\geq0$ is the length of $\row{s}$ and the 123 system has $\row{e_1}$
in the direction of $\row{s}$ by definition.
Note that we are carefully distinguishing row vectors from column vectors, as
in the scalar product $\row{s}\cdot\col{\sigma}$ of row $\row{s}$ and column
$\col{\sigma}$; of course, columns and rows are transposes of each other,
$\col{s}=\trans{\row{s}}$.
Admittedly, this distinction is somewhat pedantic, but it makes book keeping 
much easier.

Unitary transformations that affect only one of the q-bits or both q-bits 
independently are \emph{local} transformations.
Geometrically speaking, local transformations rotate $\row{\sigma}$ and
$\row{\tau}$.
Two states that can be turned into each other by a local transformation 
are \emph{locally equivalent}.
For instance, two first--q-bit states (\ref{eq:rho1}) are equivalent if their
Pauli vectors have the same length; at most the two states can differ by the
direction of $\row{e_1}$.
In other words: The difference of two equivalent states is only in the 123
bases that go with the generic form $\half(1+s\sigma_1)$.

Likewise, there are families of locally equivalent 2--q-bit states.
To decide whether two given 2--q-bit states belong to the same family, one
may put them into a generic form that is uniquely fixed by convenient 
conventions. The following set of conventions seems to be quite natural.

It all begins with recalling that the general form of a 2--q-bit state is given
by
\begin{equation}\label{eq:Rho}
\Rho=\frac{1}{4}\left(1+\row{\sigma}\cdot\col{s}+\row{t}\cdot\col{\tau}
+\row{\sigma}\cdot\dyadic{C}\cdot\col{\tau}\right)\;.  
\end{equation}
It involves the cross dyadic $\dyadic{C}$,
\begin{equation}\label{eq:cross}
  \dyadic{C}=\expect{\col{\sigma}\row{\tau}}
            =\ROW{\col{e_x},\col{e_y},\col{e_z}}
             \DYD{C_{xx} & C_{xy} & C_{xz} \\ 
                  C_{yx} & C_{yy} & C_{yz} \\ 
                  C_{zx} & C_{zy} & C_{zz}}
             \COL{\row{n_x} \\ \row{n_y} \\ \row{n_z}}\,,
\end{equation}
in addition to the Pauli vectors $\col{s}$ and $\row{t}$,
\begin{eqnarray}
 \col{s}&=&\expect{\col{\sigma}} 
        =\ROW{\col{e_x},\col{e_y},\col{e_z}}
         \COL{s_x \\ s_y \\ s_z}\,,
\nonumber\\
\row{t}&=&\expect{\row{\tau}}
       =\ROW{t_x,t_y,t_z}
        \COL{\row{n_x} \\ \row{n_y} \\ \row{n_z}}\,.
  \label{eq:s&t}
\end{eqnarray}
In a first step we bring $\dyadic{C}$ into the diagonal form
\begin{eqnarray}
&\dyadic{C}=\pm\bigl(\col{e_1}c_1\row{n_1}+\col{e_2}c_2\row{n_2}
            +\col{e_3}c_3\row{n_3}\bigr)
&
\nonumber\\[1ex]
&\mbox{for}\quad \left\{
\begin{array}{l}
\determ{\dyadic{C}}\geq0\,,\\[1ex]\determ{\dyadic{C}}<0\,,
\end{array}\right.&
\label{eq:diagC}
\end{eqnarray}
with its characteristic values ordered in accordance with
\begin{equation}\label{eq:charvals}
  c_1\geq c_2\geq c_3\geq0\,.
\end{equation}
Their squares are the eigenvalues of ~$\dyadic{C}\cdot\trans{\dyadic{C}}$
or ~$\trans{\dyadic{C}}\!\cdot\dyadic{C}$; the eigencolumns of
~$\dyadic{C}\cdot\trans{\dyadic{C}}$ constitute the orthonormal right-handed
set $\col{e_1},\col{e_2},\col{e_3}$, and the corresponding 
$\row{n_1},\row{n_2},\row{n_3}$ are eigenrows
of ~$\trans{\dyadic{C}}\!\cdot\dyadic{C}$.

Whereas the sign in (\ref{eq:diagC}) and the values of the $c_k$s are
determined by the three local invariants\footnote{\label{fn:Spur}%
We write $\Spur{\ }$ for the trace of a dyadic in order to avoid confusion
with quantum mechanical traces such as 
$C_{xy}=\olexpect{\sigma_x\tau_y}=\Tr{\sigma_x\tau_y\Rho}$.}
\begin{equation}
  \label{eq:LocInvC}
\Spur{\trans{\dyadic{C}}\!\cdot\dyadic{C}}\,,\quad 
\determ{\dyadic{C}}\,, \quad
\Spur{\left(\trans{\dyadic{C}}\!\cdot\dyadic{C}\right)^2}\,,
\end{equation}
the 123 bases are not uniquely specified by (\ref{eq:diagC}), however, because
the simultaneous replacements 
\begin{equation}
  \label{eq:notunik}
\ROW{\col{e_1},\col{e_2},\col{e_3}}\to\ROW{-\col{e_1},-\col{e_2},\col{e_3}}\,,
\quad
\COL{\row{n_1} \\ \row{n_2} \\ \row{n_3}}  \to
\COL{-\row{n_1} \\ -\row{n_2} \\ \row{n_3}}\,,  
\end{equation}
for example, do not change the right-hand side of (\ref{eq:diagC}).
The resulting freedom in choosing $\col{e_1}$, $\col{e_2}$, $\col{e_3}$
(which then fixes $\row{n_1}$, $\row{n_2}$, $\row{n_3}$ unless $c_2=0$)
is then used to enforce conventions imposed on the coefficients in
\begin{eqnarray}
 \col{s}&=&\expect{\col{\sigma}} 
        =\ROW{\col{e_1},\col{e_2},\col{e_3}}
         \COL{s_1 \\ s_2 \\ s_3}\,,\nonumber\\[1ex]
\row{t}&=&\expect{\row{\tau}}
       =\ROW{t_1,t_2,t_3}
        \COL{\row{n_1} \\ \row{n_2} \\ \row{n_3}}\,.  
  \label{eq:st123}
\end{eqnarray}
In brief terms, these conventions aim at making as many of the $s_k$s and
$t_k$s vanish as possible and to give definite signs to as many as possible of
the remaining ones. Eventually, each family is characterized by nine numbers:
the values of the three local invariants in (\ref{eq:LocInvC}), the three 
$s_k$ ($k=1,2,3$) coefficients, and the three $t_k$s, some of them equal to
zero and others with a known sign. 
Roughly speaking, of the 15 coefficients appearing in (\ref{eq:cross}) and
(\ref{eq:s&t}), six are thus used up in defining the two 123 
coordinate systems, and nine to identify the family of locally equivalent
states to which the given $\Rho$ belongs. 
Clearly, the nine family-defining parameters are invariant under local
transformations. 

Degeneracy among the characteristic values of the cross dyadic distinguishes
six classes of families,
\begin{equation}
  \label{eq:classes}
\begin{array}{r@{:\quad\mbox{class\ }}l}
c_1=c_2=c_3=0&\mathrm{A}\,,\\   
c_1=c_2=c_3>0&\mathrm{B}\,,\\   
c_1>c_2=c_3=0&\mathrm{C}\,,\\   
c_1>c_2=c_3>0&\mathrm{D}\,,\\   
c_1=c_2>c_3&\mathrm{E}\,,\\   
c_1>c_2>c_3&\mathrm{F}\,.   
\end{array}
\end{equation}
In classes A and C the $+$ sign in (\ref{eq:diagC}) applies; 
both signs can occur in classes B, D, E, and F which, therefore, 
consist of two subclasses each.

Given the local invariants of (\ref{eq:LocInvC}), we find the respective class
as follows.
First calculate the auxiliary quantities
\begin{eqnarray}
  a&=&\frac{9}{4}\Spur{\trans{\dyadic{C}}\!\cdot\dyadic{C}}
    \Spur{\left(\trans{\dyadic{C}}\!\cdot\dyadic{C}\right)^2} 
    \nonumber\\ && \mbox{} 
     -\frac{5}{4}\biggl[\Spur{\trans{\dyadic{C}}\!\cdot\dyadic{C}}\biggr]^3
     +\frac{27}{2}\biggl[\determ{\dyadic{C}}\biggr]^2\,,\nonumber\\[2ex]
  b&=&\frac{3}{2}\Spur{\left(\trans{\dyadic{C}}\!\cdot\dyadic{C}\right)^2}
    -\frac{1}{2}\biggl[\Spur{\trans{\dyadic{C}}\!\cdot\dyadic{C}}\biggr]^2\,,
\nonumber\\[-2ex] \label{eq:a&b}
\end{eqnarray}
which are subject to $a^2\leq b^3$. Then we have the classification
\begin{equation}
  \label{eq:classes2}
\begin{array}[b]{@{\mbox{class\ }}l@{\quad\mbox{if}\quad}l}
\mathrm{A}& a^2=b^3=0\AND\determ{\dyadic{C}}=0\,,\\[2ex]
\mathrm{B}& a^2=b^3=0\AND\determ{\dyadic{C}}\neq0\,,\\[2ex]
\mathrm{C}& a^2=b^3>0\begin{array}[t]{@{}l}
\AND a>0 \\ \AND \determ{\dyadic{C}}=0\,,\end{array}\\[2ex]
\mathrm{D}& a^2=b^3>0\begin{array}[t]{@{}l}
\AND a>0\\  \AND \determ{\dyadic{C}}\neq0\,,\end{array}\\[2ex]
\mathrm{E}& a^2=b^3>0\AND a<0\,,\\[2ex]
\mathrm{F}& a^2<b^3\,.
\end{array}
\end{equation}
The generic forms for the various classes are as follows.

\textbf{Class A:} 
Since $\dyadic{C}=0$ here, we can choose the two sets of 123 coordinates
independently, and $s_1\geq0$, $s_2=s_3=0$ as well as $t_1\geq0$, $t_2=t_3=0$
specify the conventions.
This class consists of a two-parametric set of families of the generic form
\begin{equation}
  \label{eq:genA}
  \Rho=\frac{1}{4}\bigl(1+s\sigma_1+t\tau_1\bigr)\WITH s\geq0\,,\ t\geq0\,.
\end{equation}
For $s=0$, $t=0$ we have the chaotic state $\chaos=\frac{1}{4}$ which forms a
single-state family all by itself.

\textbf{Class B:}
Here we can choose $\col{e_1},\col{e_2},\col{e_3}$ freely and the conventional
choice is specified by
\begin{eqnarray}
 && \col{s}=\col{e_1}s\,,\quad\row{t}=t_1\row{n_1}+t_3\row{n_3}
  \nonumber\\[1ex]
&\textrm{with}&\left\{
    \begin{array}{c}
    s>0\AND t_3\geq0\\ \OR \\ s=0 \AND t_1=t\geq0\,,\ t_3=0
    \end{array}\right\}\,.
  \label{eq:genB}
\end{eqnarray}
Each subclass [$\pm$ in (\ref{eq:diagC})] consists of four-parametric sets of
families.
In passing we note that the so-called Werner states constitute the two class-B
families with $s=0$ and $t=0$.

\textbf{Class C:}
Here the replacement (\ref{eq:notunik}) is used to enforce $s_1\geq0$ or
$t_1\geq0$ if $s_1=0$. Then $s_2=0,s_3\geq0$ and $t_2=0,t_3\geq0$ are achieved
by suitable rotations of $\col{e_2},\col{e_3}$ and, independently, of
$\row{n_2},\row{n_3}$. In summary, this establishes
\begin{eqnarray}
&&
\col{s}=\col{e_1}s_1+\col{e_3}s_3\,,\quad\row{t}=t_1\row{n_1}+t_3\row{n_3}\,,
\nonumber\\
&&\dyadic{C}=\col{e_1}c_1\row{n_1}\nonumber\\[2ex]
&\textrm{with}&\left\{\begin{array}{c}
s_1\geq0\\ \OR\\ s_1=0\AND t_1\geq0
\end{array}\right\}
\nonumber\\[1ex]&
\textrm{and}& s_3\geq0\,,\ t_3\geq0
  \label{eq:genC}
\end{eqnarray}
for the five-parametric sets of families.

\textbf{Class D:}
In distinction from class C, the rotations in the 23 sectors are not
independent here. 
Thus we get
\begin{eqnarray}
&&
\col{s}=\col{e_1}s_1+\col{e_3}s_3\,,\quad
\row{t}=t_1\row{n_1}+t_2\row{n_2}+t_3\row{n_3}\,,
\nonumber\\
&&\dyadic{C}=\pm\bigl(\col{e_1}c_1\row{n_1}
                        +\col{e_2}c_2\row{n_2}
                        +\col{e_3}c_2\row{n_3}\bigr)
\nonumber\\[2ex]
&\textrm{with}&\left\{\begin{array}{c}
s_1\geq0\\ \OR\\ s_1=0\AND t_1\geq0
\end{array}\right\} \nonumber\\[1ex] 
&\textrm{and}&\left\{\begin{array}{c} 
s_3\geq0\AND t_2\geq0\\ \OR \\ s_3=0\AND t_2=0\,,\ t_3\geq0
\end{array}\right\}\,.
  \label{eq:genD}
\end{eqnarray}
Each subclass contains seven-parametric sets of families.

\textbf{Class E:}
This class is very similar to class D, but now the degeneracy is in the 12
sector, and so we have
\begin{eqnarray}
&&
\col{s}=\col{e_1}s_1+\col{e_3}s_3\,,\quad
\row{t}=t_1\row{n_1}+t_2\row{n_2}+t_3\row{n_3}\,,
\nonumber\\ &&\dyadic{C}=\pm\bigl(\col{e_1}c_1\row{n_1}
                        +\col{e_2}c_1\row{n_2}
                        +\col{e_3}c_3\row{n_3}\bigr)
\nonumber\\[2ex]
&\textrm{with}&
\left\{\begin{array}{c} 
s_1\geq0\AND t_2\geq0\\ \OR \\ s_1=0\AND t_1\geq0\,,\ t_2=0
\end{array}\right\}
\nonumber\\[1ex] &\textrm{and}&
\left\{\begin{array}{c}
s_3\geq0\\ \OR\\ s_3=0\AND t_3\geq0
\end{array}\right\}\,.
  \label{eq:genE}
\end{eqnarray}
Here, too, each subclass is made up of seven-parametric sets of families.

\textbf{Class F:}
The lack of degeneracy limits changes of the 123 bases to discrete
$180^{\circ}$ rotations as in (\ref{eq:notunik}) where the rotation
is around the 3rd axes. The generic form is then defined by that choice of 123
coordinates for which as many as possible of the coefficients $s_1$, $t_1$,
$s_2$, $t_2$, $s_3$, $t_3$, are non-negative (in this order, say). Here we
get, in each subclass, sets of families specified by the full number of nine
parameters, of which five or more are non-negative.

Arbitrary local unitary transformations turn members of a family into other
members of the same family --- this, we recall, is the defining property of a
family of locally equivalent states.
It is possible that some local transformations have no effect at all, as
exemplified by $U_{\mathrm{loc}}=\exp(i\varphi\sigma_1+i\phi\tau_1)$ acting on
the class-A state (\ref{eq:genA}).
Therefore, some families are larger than others, and determining a family's
size is a problem of considerable interest.
Recent progress on this front is reported by Ku\'s and \.Zyczkowski
\cite{K+Z}.

We close this section with a single example. 
Pure states are of the generic form
\begin{eqnarray}
&  \pure=\frac{1}{4}\left(1+p\sigma_1-p\tau_1
         -\sigma_1\tau_1-q\sigma_2\tau_2-q\sigma_3\tau_3\right)  &
\nonumber\\
&\WITH  0\leq p\leq1\,,\ q=\sqrt{1-p^2}\geq0\,.&
  \label{eq:pure}
\end{eqnarray}
One verifies easily the purity condition
\begin{equation}
  \label{eq:purecond}
  \pure\left(1-\pure\right)=0\,.
\end{equation}
For $p=0,q=1$ we have the family of Bell states,
\begin{equation}
  \label{eq:Bell}
  \Bell=\frac{1}{4}\left(1-\sigma_1\tau_1-\sigma_2\tau_2-\sigma_3\tau_3\right)
\,,
\end{equation}
which is in class B; 
the $p=1,q=0$ family consists of the product states
$\half(1+\sigma_1)\half(1-\tau_1)$ and is in class C;
and the $0<p<1$ families belong to class D.
These families are of different sizes \cite{K+Z}:
three-dimensional, four-dimensional, and five-dimensional, respectively.

\section{Positivity and separability}

An arbitrary choice for the (real) coefficients in (\ref{eq:Pauli-ops,xyz})
and (\ref{eq:cross}) or, equivalently, of the nine family-defining parameters
plus the 123 coordinate systems specifies a hermitian $\Rho$ of unit trace,
but its positivity must be ensured by imposing restrictions on the Pauli
vectors $\col{s}$, $\row{t}$, and the cross dyadic $\dyadic{C}$.
It is expedient to switch the emphasis from $\Rho$ to the traceless operator
$K$ introduced by
\begin{equation}
  \label{eq:Kdef}
  \Rho=\frac{1}{4}(1-K)\,,\qquad K=1-4\Rho\,,
\end{equation}
so that $\Rho\geq0$ requires
\begin{equation}
  \label{eq:posit}
  K\leq1\,.
\end{equation}
Convex sums of two states are weighted sums of their $K$s.
Admixing $\chaos$ to a given $\Rho$ amounts to multiplying its $K$ by a factor.

One could, of course, check the positivity criterion (\ref{eq:posit}) by
calculating the eigenvalues of $K$.
For such purposes, it is often very convenient to use a $4\times4$-matrix
representation in which $\sigma_{1,2,3}$ and $\tau_{1,2,3}$ have imaginary
antisymmetric matrices,  
\begin{eqnarray}
\lefteqn{\row{\sigma}\cdot\col{s}+\row{t}\cdot\col{\tau}}
\nonumber\\
&&\widehat{=}\mbox{\small$\displaystyle
\left(\begin{array}{c@{\enskip}c@{\enskip}c@{\enskip}c}
0 & -i(s_1+t_1) & +i(s_2+t_2) & -i(s_3-t_3)\\
+i(s_1+t_1) &0 & +i(s_3+t_3) & +i(s_2-t_2)\\
-i(s_2+t_2) & -i(s_3+t_3) & 0 & +i(s_1-t_1)\\
+i(s_3-t_3) & -i(s_2-t_2) & -i(s_1-t_1) & 0
\end{array}\right)$}\,,\nonumber\\
  \label{eq:sigtau-matr}
\end{eqnarray}
and products $\sigma_j\tau_k$ ($j,k=1,2,3$) have real symmetric matrices, in
particular
\begin{eqnarray}
\lefteqn{ \displaystyle
 \row{\sigma}\cdot\dyadic{C}\cdot\col{t}=\pm\sum_{k=1}^3\sigma_kc_k\tau_k}
\nonumber\\[1ex] &&\widehat{=}\pm\mbox{\small$
\left(\begin{array}{c@{\enskip}c@{\enskip}c@{\enskip}c}
c_1+c_2-c_3 &0&0&0\\
0& c_1-c_2+c_3 &0&0\\
0&0& -c_1+c_2+c_3 &0\\
0&0&0& -c_1-c_2-c_3
\end{array}\right)$}\nonumber\\
  \label{eq:sCt-matr}
\end{eqnarray}
is diagonal.
But precise knowledge of the actual eigenvalues of $K$ is not needed if we
only want to verify (\ref{eq:posit}). 

Since $K$ is traceless, its eigenvalues $\kappa_j$ ($j=1,2,3,4$) have a
vanishing sum and solve a quartic equation without a cubic term,
\begin{equation}
  \label{eq:quartic}
  \kappa^4-A_2\kappa^2+A_1\kappa-A_0=0\,,
\end{equation}
where
\begin{eqnarray}
  A_2&=&\frac{1}{2}\Tr{K^2}\,,\nonumber\\[1ex]
  A_1&=&-\frac{1}{3}\Tr{K^3}\,,\nonumber\\[1ex]
  A_0&=&\frac{1}{4}\Tr{K^4}-\frac{1}{8}\left[\Tr{K^2}\right]^2\,.
  \label{eq:As1}
\end{eqnarray}
These three numbers are invariant under arbitrary (local or not) unitary
transformations; they are three independent \emph{global} invariants of the
given $\Rho$.
Expressed in terms of $\col{s}$, $\row{t}$, and $\dyadic{C}$ they read
\begin{eqnarray}
A_2&=&2\Spur{\trans{\dyadic{C}}\!\cdot\dyadic{C}}+2(s^2+t^2)\,,\nonumber\\
A_1&=&8\determ{\dyadic{C}}-8\row{s}\cdot\dyadic{C}\cdot\col{t}\,,\nonumber\\
A_0&=&-\left(\half A_2\right)^2
      +2\left[\Spur{\trans{\dyadic{C}}\!\cdot\dyadic{C}}\right]^2
      +4s^2t^2\nonumber\\ &&
      +4\row{s}\cdot\dyadic{C}\cdot\trans{\dyadic{C}}\cdot\col{s}
      +4\row{t}\cdot\trans{\dyadic{C}}\cdot\dyadic{C}\cdot\col{t}\nonumber\\ &&
      +8\determ{\dyadic{E}}-8\determ{\dyadic{C}}
\label{eq:As2}
\end{eqnarray}
where
\begin{equation}
  \label{eq:defE}
  \dyadic{E}=\dyadic{C}-\col{s}\row{t}
\end{equation}
is the entanglement dyadic.

As we see, the traces of (\ref{eq:As1}) involve nine different local
polynomial invariants of $\col{s}$, $\row{t}$, and $\dyadic{C}$, and it is
clear that their values are determined by the nine family-specifying
parameters of classes A,\dots, F.
Suggestive as it is, the converse is not true,%
\footnote{Therefore, the assertion ``All other local invariants \dots''
shortly after (19) in \cite{EngMet} is false.}\  
as can be demonstrated by a counter example. Consider, for instance, the two
states 
\begin{eqnarray}
\Rho_1&=&\frac{1}{4}\left(1+\frac{1}{4}\sigma_1+\half\sigma_3
          +\half\sigma_1\tau_1+\frac{1}{4}\sigma_2\tau_2\right)\,,
\nonumber\\
\Rho_2&=&\frac{1}{4}\left(1+\half\sigma_2+\frac{1}{4}\sigma_3
          +\half\sigma_1\tau_1+\frac{1}{4}\sigma_2\tau_2\right)\,,
  \label{eq:counterexpl}  
\end{eqnarray}
which belong to two different class-F families, but all terms in
(\ref{eq:As2}) are the same for $\Rho_1$ and $\Rho_2$.

Whereas the nine polynomial invariants of (\ref{eq:As2}) do not always suffice
to determine the values of all local invariants, the nine parameters that
specify the family certainly do.
They are, however, not given by (traces of) polynomials of  $\col{s}$,
$\row{t}$, and $\dyadic{C}$.
According to Makhlin \cite{Makhlin}, there are 18 polynomial invariants whose
values uniquely characterize the family in question (actually, of nine of them
only the sign matters).
In addition to the nine invariants in (\ref{eq:As2}), which exhaust the
polynomials of degree 4 or lower, there are nine invariants of higher degree
in Makhlin's set, which do not enter the three global invariants $A_0$, $A_1$,
$A_2$. 

All solutions of the quartic equation (\ref{eq:quartic}) are real by
construction --- it is, after all, the characteristic polynomial of a
hermitian operator.
Then, if all solutions are in the range $\kappa\leq1$, this polynomial and its
derivatives must be non-negative for $\kappa\geq1$.
Consequently, the positivity requirement (\ref{eq:posit}) implies
\begin{equation}
  \label{eq:ineqs}
  A_2-A_1+A_0\leq1\,,\quad 2A_2-A_1\leq4\,,\quad A_2\leq6\,.
\end{equation}
The converse is also true: If these three inequalities are obeyed, the four
real solutions of (\ref{eq:quartic}) are in the range $\kappa\leq1$, so that
$K\leq1$ and $\Rho\geq0$.
In other words, the restrictions on $\col{s}$, $\row{t}$, and $\dyadic{C}$
alluded to at the beginning of this section are just the inequalities
(\ref{eq:ineqs}).

Although the equivalence of (\ref{eq:posit}) and (\ref{eq:ineqs}) is rather
obvious, a clear-cut demonstration of the case could be of interest to some
readers. We give one in the Appendix.

If the entanglement dyadic $\dyadic{E}$ vanishes, the state in question is of
product form,
\begin{equation}
  \label{eq:prodform}
  \Rho=\half\left(1+\row{\sigma}\cdot\col{s}\right)\,
       \half\left(1+\row{t}\cdot\col{\tau}\right)\,,
\end{equation}
so that results of measurements on the first q-bit show no correlations
whatsoever with measurement results concerning the second q-bit.
Under these circumstances the 2--q-bit system is \emph{not entangled}.
Entangled q-bit pairs, $\dyadic{E}\neq0$, may be in a mixed state blended from
disentangled ingredients,
\begin{eqnarray}
&\displaystyle
 \Rho=\sum_n w_n\half\left(1+\row{\sigma}\cdot\col{s_n}\right)\,
       \half\left(1+\row{t_n}\cdot\col{\tau}\right)&\nonumber\\
&\displaystyle \WITH  w_n>0\,,\ \sum_n w_n=1\,;&
  \label{eq:classmix}
\end{eqnarray}
then all correlations found in the measurement data can be understood 
classically.
States of this kind are called \emph{separable}.
As an elementary example, consider the pure states (\ref{eq:pure}):
For $p=1$ they are not entangled and therefore separable,
for $p<1$ they are entangled and not separable.

Correlations of a genuine quantum character require a non-separable state
$\Rho$.
According to a criterion that we owe to Peres \cite{Per} as well as  
M., P., and R. Horodecki \cite{MPRHor}, a given state $\Rho$ is separable if
\begin{eqnarray}
  \tRho&=&\frac{1}{4}\left(1-\row{\sigma}\cdot\col{s}
                          +\row{t}\cdot\col{\tau}
                          -\row{\sigma}\cdot\dyadic{C}\cdot\col{\tau}\right)
\nonumber\\
       &=&\half\left(\row{\sigma}\cdot\Rho\col{\sigma}-\Rho\right)
  \label{eq:tRho}
\end{eqnarray}
is positive and only then.
Let's call $\tRho$ the \PHHH\ transform of $\Rho$; it is unitarily equivalent to
the partial transpose originally considered by Peres and the Horodeckis.

The positivity of $\tRho$, or
\begin{equation}
  \label{eq:separ}
  1-4\tRho=\tK=\half\left(\row{\sigma}\cdot K\col{\sigma}-K\right)\leq1\,,
\end{equation}
can be checked analogously to the positivity of $\Rho$.
Now, the quartic equation solved by the eigenvalues
$\tkappa_1,\dots,\tkappa_4$ of $\tK$ is
\begin{eqnarray}
&&  \tkappa^4-A_2\tkappa^2+\left(A_1+16\determ{\dyadic{C}}\right)\tkappa
\nonumber\\ &&\quad\mbox{}          
-\left(A_0-16\determ{\dyadic{E}}+16\determ{\dyadic{C}}\right)=0\,,  
  \label{eq:tquartic}
\end{eqnarray}
so that
\begin{eqnarray}
  A_2-A_1+A_0&\leq&1+16\determ{\dyadic{E}}\,,\nonumber\\[1ex]
  2A_2-A_1&\leq&4+16\determ{\dyadic{C}}\,,\nonumber\\[1ex] 
  A_2&\leq&6\,.
  \label{eq:tineqs}
\end{eqnarray}
are equivalent to (\ref{eq:separ});
the third is always obeyed by a positive $\Rho$.
So, a non-separable state must violate either the first or the second
inequality, or both.
The equal sign holds in the first inequality, if the \PHHH\ transform of the
given $\Rho$ has a zero eigenvalue; the first and the second are
equalities, if the \PHHH\ transform has two zero eigenvalues.
Accordingly, the $\tRho$ of a non-separable $\Rho$ can at most have one zero
eigenvalue and thus must be of rank 3 or~4.
While we are at it, let us also mention that the \PHHH\ transform of any
state $\Rho$ can have at most a single negative eigenvalue 
[see below at Eq.\ (\ref{eq:pseudo})]. 

Thus the separability of a given $\Rho$ is checked as easily as its
positivity. Neither test requires actual knowledge of the solutions of
(\ref{eq:quartic}) or (\ref{eq:tquartic}).
They could, of course, be stated analytically but these explicit 
expressions are 
not very transparent unless special relations exist among the coefficients of
the quartic equations. 

As an immediate implication of the \PHHH\ criterion, in the form of the
inequalities (\ref{eq:tineqs}), we note that a state $\Rho$ with
$\determ{\dyadic{C}}\geq0$ and $\determ{\dyadic{E}}\geq0$ is surely
separable.
Therefore, for example, all states in classes A and C are separable.

\section{Lewenstein-Sanpera decompositions}
\label{sec:LSD}
As Lewenstein and Sanpera observed \cite{LS}, any 2--q-bit state $\Rho$ can be
written as a convex sum of a separable state and a pure state,
\begin{equation}
  \label{eq:LSD}
  \Rho=\lambda\sep+(1-\lambda)\pure \WITH 0\leq\lambda\leq1\,.
\end{equation}
Rare exceptions aside, the \emph{Lewenstein-Sanpera decomposition} (LSD) 
of a given (non-separable) $\Rho$ is not unique, 
there is usually a continuum of LSDs to choose from.
Among them is the \emph{optimal LSD}, the one with the largest value of
$\lambda$,
\begin{equation}
  \label{eq:optLSD}
  \Rho=\DoS\sepopt+(1-\DoS)\pureopt \WITH
  \DoS=\max\bigl\{\lambda\bigr\}\,,
\end{equation}
and we call $\DoS$, the maximal $\lambda$ value, 
the \emph{degree of separability} of $\Rho$. 
Without presently attempting to be precise about this matter, we repeat the
remark in \cite{EngMet} that ``a state $\Rho$ is the more useful for quantum
communication purposes, the smaller its degree of separability.''

The spectral decomposition of the \PHHH\ transform of a pure state
(\ref{eq:pure}) is of the generic form
\begin{eqnarray}
\tpure&=&\frac{1}{4}\left(1-p\sigma_1-p\tau_1
                          +\sigma_1\tau_1
                          +q\sigma_2\tau_2
                          +q\sigma_3\tau_3\right)\nonumber\\[1ex]
&=&\frac{1+p}{2}\pure^{(1)}+\frac{1-p}{2}\pure^{(2)}
      +\frac{q}{2}\pure^{(3)}-\frac{q}{2}\pure^{(4)}
\nonumber\\[-1ex]  \label{eq:pureP3H}
\end{eqnarray}
with
\begin{equation}
  \label{eq:pureP3H.a}
\left.\begin{array}{l}
\pure^{(1)} \\[1ex] \pure^{(2)}
\end{array}\right\}
=\frac{1}{4}\left(1\mp\sigma_1\mp\tau_1+\sigma_1\tau_1\right)\,,
\end{equation}
which are pure states of the separable class-C kind, and
\begin{equation}
  \label{eq:pureP3H.b}
\left.\begin{array}{l}
\pure^{(3)} \\[1ex] \pure^{(4)}
\end{array}\right\}
=\frac{1}{4}\left(1-\sigma_1\tau_1\pm\sigma_2\tau_2\pm\sigma_3\tau_3\right)\,,
\end{equation}
which are Bell states (non-separable, class B).
Therefore, the \PHHH\ transform $\tRho$ of any 2--q-bit state $\Rho$ can be 
written as
\begin{equation}
  \label{eq:pseudo}
  \tRho=(1+x)\Rho'-x\Bell\,,\quad 0\leq x\leq\half(1-\DoS)
\end{equation}
with some state $\Rho'$ and a Bell state $\Bell$.
As a consequence, $\tRho$ can have at most one negative eigenvalue, so that
only one solution of (\ref{eq:tquartic}) can be in the range $\tkappa>1$, as
noted above.

Since $\Rho'$ is a mixture of four or fewer pure states, (\ref{eq:pseudo})
shows that the \PHHH\ transform of a non-separable state is a pseudo-mixture of
up to five pure states with one negative weight only, carried by a Bell state.
There is a very similar observation by Sanpera, Tarrach, and Vidal \cite{STV}
about $\Rho$ itself: It can always be presented as a pseudo-mixture of four
or five separable pure states;
as an immediate consequence its \PHHH\ transform is also such a pseudo-mixture.

The optimal LSD (\ref{eq:optLSD}) has a number of properties that help in
decomposing given states in the optimal way.
Let's briefly consider some particularly important ones.

\textbf{Existence:}
The degree of separability $\DoS$ is really the maximum of all possible
$\lambda$ values in (\ref{eq:LSD}), not just their supremum, because the
subset of separable states is compact.
Therefore, a LSD with $\lambda=\DoS$ does exist.

\textbf{Uniqueness:}
If we have two different LSDs with the same non-zero value of $\lambda$, 
their symmetric  convex sum also equals the given $\Rho$.
It contains the convex sum of the two different $\sep$s, which is separable,
and the convex sum of the two $\pure$s, which has LSDs of its own.
Either one of them contains a separable part, so that we get a new LSD of the
$\Rho$ in question with a larger $\lambda$ value.
Consequently, the common $\lambda$ of the original two LSDs is not maximal,
and it follows that the optimal LSD is unique.

This does not imply that one can always find another LSD with the same
$\lambda$ value if $\lambda<\DoS$.
There are $\Rho$s with a continuum of LSDs in which each value of $\lambda$
occurs only once.%
\footnote{Therefore, the `only' is too strong in the assertion ``Only $\sep$
and $\pure$ \dots'' after (15) in \cite{EngMet}.}\ 
Examples are the rank-2 states of (56) in \cite{EngMet} that obey inequality
(61) in \cite{EngMet}.

\textbf{$\sepopt$ is barely separable:}
Consider the optimal LSD of some non-separable $\Rho$ and a parameter
$\epsilon$ in the range $0<\epsilon\leq1-\DoS$. 
In
\begin{eqnarray}
  \Rho&=&(\DoS+\epsilon)\left[\frac{\DoS}{\DoS+\epsilon}\sepopt
                         +\frac{\epsilon}{\DoS+\epsilon}\pureopt\right]
\nonumber\\   
  && +(1-\DoS-\epsilon)\pureopt
  \label{eq:baresep'}
\end{eqnarray}
the convex sum in square brackets is surely non-negative, but cannot be
separable.
Because, if it were, we would have found a LSD with $\lambda>\DoS$.
Thus the \PHHH\ transform of $\bigl[\cdots\bigr]$ has a negative eigenvalue
for $\epsilon>0$, but none for $\epsilon=0$.
Since the eigenvalues are continuous functions of~$\epsilon$, the \PHHH\
transform of $\sepopt$ must have at least one zero eigenvalue.
Formally,
\begin{equation}
  \label{eq:baresep}
  \tsepopt\geq0\quad\mbox{but not}\quad\tsepopt>0\,;
\end{equation}
for $\sepopt$, the equal sign holds in the first inequality of
(\ref{eq:tineqs}). 
A useful terminology calls $\sepopt$ \emph{barely separable} with respect to 
$\pureopt$.

When searching for the optimal LSD of a given $\Rho$ it is, therefore,
sufficient to consider LSDs with $\sep$s that are barely separable with
respect to the $\pure$ with which they are paired in (\ref{eq:LSD}).
If the $\sep$ of some LSD does not have this property, one adds the appropriate
amount of the respective $\pure$ to it (in the sense of a convex sum, of course)
and gets a barely separable $\sep$.

\textbf{Local invariance is passed on:}
Suppose that the $\Rho$ considered is invariant under some local unitary
transformation,
\begin{equation}
  \label{eq:Uloc}
  U_{\mathrm{loc}}^{\dagger}\Rho U_{\mathrm{loc}}^{\phantom{\dagger}}=\Rho\,.
\end{equation}
Then its $\sepopt$ and $\pureopt$ must be invariant under this local
transformation as well.
Otherwise we could apply it to the optimal LSD and get another LSD with the
same $\lambda$ value, in conflict with the uniqueness of the optimal LSD.
This argument builds on the elementary observation that local transformations
do not affect the purity and separability of a state.

The limitations resulting from this ``inheritance of local invariance'' can
facilitate the search for the optimal LSD substantially.
The optimal decompositions of the (generalized) Werner states reported in
\cite{EngMet} were found this way.

\textbf{Swapping invariance is passed on:}
Similarly one finds that the $\sepopt$ and $\pureopt$ of a $\Rho$ that is
invariant under the swapping transformation
\begin{equation}
  \label{eq:swap}
  \sigma_k\leftrightarrow\tau_k\quad\textrm{for}\quad k=1,2,3
\end{equation}
must be invariant themselves because swapping does not affect the separability
or the purity of a state.
Clearly, this swapping invariance is only possible if the Pauli vectors
$\col{s}$ and $\row{t}$ are of equal length.

\textbf{Orthogonality is passed on:}
If the $\Rho$ in question is orthogonal to a certain other state
$\Rho_{\perp}$,
\begin{equation}
  \label{eq:orth}
  \Tr{\Rho\Rho_{\perp}}=0\,,
\end{equation}
then the $\sep$s and $\pure$s of all LSDs of $\Rho$ are also orthogonal to
$\Rho_{\perp}$ because both traces in
\begin{equation}
  \label{eq:orth'}
  0=\lambda\Tr{\sep\Rho_{\perp}}+(1-\lambda)\Tr{\pure\Rho_{\perp}}
\end{equation}
must be non-negative, so both must vanish.
In particular, the $\sepopt$ and $\pureopt$ of $\Rho$ must have this
orthogonality property.  
The optimal LSDs of rank-2 states reported in \cite{EngMet} were found by
exploiting this ``inheritance of orthogonality.''

\section{Degree of separability and concurrence}
\label{sec:concur}
In their studies of what they call ``entanglement of formation,''
Hill and Wootters \cite{HW,Woo} consider
\begin{eqnarray}
  \bRho&=&\frac{1}{4}\left(1-\row{\sigma}\cdot\col{s}-\row{t}\cdot\col{\tau}
        +\row{\sigma}\cdot\dyadic{C}\cdot\col{\tau}\right)
\nonumber\\
  &=&\half\left(\row{\tau}\cdot\tRho\col{\tau}-\tRho\right)\,;
  \label{eq:HW}
\end{eqnarray}
let's call it the \HW\ transform of $\Rho$.
Since the replacement 
$\bigl(\col{s},\row{t},\dyadic{C}\bigr)%
\to\bigl(-\col{s},-\row{t},\dyadic{C}\bigr)$
changes none of the local invariants in (\ref{eq:As2}), 
$\bRho$ has the same eigenvalues
as $\Rho$ and, therefore, $\bRho$ is unitarily equivalent to $\Rho$.
Equally well we could argue that the matrix representations of $\Rho$ and
$\bRho$, composed of the ingredients in (\ref{eq:sigtau-matr}) and
(\ref{eq:sCt-matr}), are complex conjugates or transposes of each other, and
so they must have the same real eigenvalues.
Note that a 2--q-bit state and its \HW\ transform are always in the same class 
of states but they may or may not belong to the same family; 
their unitary equivalence may be local or global.

Hill and Wootters use the \HW\ transform to introduce the \emph{concurrence} 
$\WHC$ of $\Rho$. It is given by
\begin{equation}
  \label{eq:conc}
  \WHC=\max\left\{0,2\max_k\left\{r_k\right\}
        -\sum_k r_k\right\}\,,
\end{equation}
where $r_1,r_2,r_3,r_4$ are the four non-negative eigenvalues of
\begin{equation}
  \label{eq:HW.1}
 \magn{\textstyle\sqrt{\Rho\vphantom{\bRho}}\,\sqrt{\bRho}}
={\textstyle\sqrt{\sqrt{\bRho}\,\Rho\sqrt{\bRho}\,}}\,. 
\end{equation}
The roles of $\Rho$ and $\bRho$ can be interchanged in this definition of
$\WHC$; thus the concurrence of $\bRho$ is equal to the concurrence of $\Rho$.
For instance, the concurrence of a pure state (\ref{eq:pure}) is $q$.

Separable states ($\DoS=1$) have $\WHC=0$ and non-separ\-able states ($\DoS<1$)
have $\WHC>0$.
This suggests that there might be a close relation between the degree of
separability and the concurrence. 
Indeed, $\DoS+\WHC=1$ holds if $\col{s}=0$ and $\row{t}=0$ --- such $\Rho$s
are generalized Werner states of the first kind in the terminology of
\cite{EngMet} --- but more generally we find
\begin{equation}
  \label{eq:S+C}
   0<\DoS+\WHC\leq1\,.
\end{equation}
Pure states (\ref{eq:pure}) have $\DoS+\WHC=1$ if $q=0$ and $\DoS+\WHC=q$ if 
$q>0$.  
A set of non-pure states exploring the whole range of (\ref{eq:S+C}) is given
by the rank-2 states  
\begin{equation}
  \Rho=\frac{1}{4}\Bigl(1+(\sigma_3+x\tau_3)\sin\theta
         +(\sigma_1\tau_1-x\sigma_2\tau_2)\cos\theta+x\sigma_3\tau_3\Bigr)
  \label{eq:HW.rk2}
\end{equation}
with $-1< x<1$, for which
\begin{eqnarray}
  \label{eq:HW.rk2'}
\DoS=\left\{\begin{array}{c@{\ \textrm{if}\ }l}
     1 & \cos\theta=0\\
     1-\magn{x} & \cos\theta\neq0
\end{array}\right\}
\,,\quad\WHC=\magn{x\cos\theta}\,.
\end{eqnarray}
We surmise that (\ref{eq:S+C}) is obeyed by all 2--q-bit states.

\section*{Appendix}
We write the four solutions of (\ref{eq:quartic}) in terms of three
parameters,
\begin{eqnarray}
\left.\begin{array}{l} \kappa_1\\[0ex]\kappa_2 \end{array}\right\}
&=&\pm(\lambda_1-\lambda_2)-\lambda_3\,,\nonumber\\
\left.\begin{array}{l} \kappa_3\\[0ex]\kappa_4 \end{array}\right\}
&=&\mp(\lambda_1+\lambda_2)+\lambda_3\,,
  \label{eq:App1}
\end{eqnarray}
thereby taking care of $\Tr{K}=\kappa_1+\kappa_2+\kappa_3+\kappa_4=0$.
The coefficients in (\ref{eq:quartic}) are then given by
\begin{eqnarray}
A_2&=&\half\left(\kappa_1^2+\kappa_2^2+\kappa_3^2+\kappa_4^2\right)
\nonumber\\
      &=&2\left(\lambda_1^2+\lambda_2^2+\lambda_3^2\right)\,,
\nonumber\\[1ex]
A_1&=&\kappa_1\kappa_2(\kappa_1+\kappa_2)+\kappa_3\kappa_4(\kappa_3+\kappa_4)
\nonumber\\
      &=&-8\lambda_1\lambda_2\lambda_3\,,\nonumber\\[1ex]
A_0&=&-\kappa_1\kappa_2\kappa_3\kappa_4
    =2\left(\lambda_1^2\lambda_2^2
         +\lambda_2^2\lambda_3^2+\lambda_3^2\lambda_1^2\right)
\nonumber\\
&&\hphantom{-\kappa_1\kappa_2\kappa_3\kappa_4=}
  -\left(\lambda_1^4+\lambda_2^4+\lambda_3^4\right)\,.
  \label{eq:App2}
\end{eqnarray}
Now look at the second inequality in (\ref{eq:ineqs}). It says
\begin{equation}
  \label{eq:App3}
  \lambda_1^2+\lambda_2^2+\lambda_3^2+2\lambda_1\lambda_2\lambda_3\leq1
\end{equation}
or, singling out $\lambda_3$,
\begin{equation}
  \label{eq:App4}
  \left(\lambda_1\lambda_2+\lambda_3\right)^2
\leq\left(1-\lambda_1^2\right)\left(1-\lambda_2^2\right)\,,
\end{equation}
and cyclic permutations produce two analogous statements in which $\lambda_1$
and $\lambda_2$ are privileged.
Since the left-hand sides of these three equations are non-negative, it
follows that all $\lambda_j^2$ exceed unity if one of them does.
Combined with the third inequality in (\ref{eq:ineqs}), here reading
\begin{equation}
  \label{eq:App5}
  \lambda_1^2+\lambda_2^2+\lambda_3^2\leq3\,,
\end{equation}
this implies
\begin{equation}
  \label{eq:App6}
  \lambda_1^2\leq1\,,\qquad
  \lambda_2^2\leq1\,,\qquad
  \lambda_3^2\leq1\,.
\end{equation}
In conjunction with
\begin{eqnarray}
\lambda_1&=&\half(\kappa_1+\kappa_4)=-\half(\kappa_2+\kappa_3)\,,\nonumber\\
\lambda_2&=&\half(\kappa_2+\kappa_4)=-\half(\kappa_1+\kappa_3)\,,\nonumber\\
\lambda_3&=&\half(\kappa_3+\kappa_4)=-\half(\kappa_1+\kappa_2)
  \label{eq:App7}
\end{eqnarray}
this means that at most one of the four $\kappa$s can be larger than $1$,
and that then the other three must be less than~$1$.
But the first inequality in (\ref{eq:ineqs}),
\begin{equation}
  \label{eq:App8}
  1-A_2+A_1-A_0=(1-\kappa_1)(1-\kappa_2)(1-\kappa_3)(1-\kappa_4)\geq0\,,
\end{equation}
excludes this possibility because one negative factor and three positive
factors would yield a negative product.
Therefore, all three inequalities (\ref{eq:ineqs}) can only be obeyed if all
four $\kappa$s are less than~$1$.
In other words: (\ref{eq:ineqs}) implies (\ref{eq:posit}) indeed.

Note that this reasoning is only valid if one knows, as we do, that all
solutions of the quartic equation (\ref{eq:quartic}) are real.
This property itself is not guaranteed by the inequalities (\ref{eq:ineqs}),
as shown by $A_2=2$, $A_1=0$, $A_0=-2$ when $(\kappa^2-1)^2+1=0$.

\section*{Acknowledgments}
BGE thanks Marek Ku\'s, Maciej Lewenstein, and Karol \.Zycz\-kowski for
particularly helpful discussions.
NM would like to thank the Egyptian government for granting a fellowship.

\vfill

\setlength{\fboxsep}{10pt}
\begin{center}
\framebox{\parbox{0.75\columnwidth}{%
\begin{center}
This paper has been submitted to\\
Applied Physics B\\      
as a contribution to the\\ 
Proceedings of the DPG Spring Meeting,\\ 
held in Bonn, April 2000.
\end{center}}}      
\end{center}

\vfill

\end{document}